\documentclass[twocolumn,showpacs,prc,floats,amsmath,amssymb]{revtex4}
\usepackage{graphicx}% Include figure files
\usepackage{dcolumn}% Align table columns on decimal point
\usepackage{color}
\usepackage{amssymb}
\usepackage{amsfonts}
\usepackage{amsmath}
\usepackage{revsymb}
\usepackage{epsfig}
\usepackage{graphics}
\newcommand{\be}{\begin{equation}}
\newcommand{\ee}{\end{equation}}
\newcommand{\ba}{\begin{eqnarray}}
\newcommand{\ea}{\end{eqnarray}}
\newcommand{\sba}{\begin{subequations}}
\newcommand{\sea}{\end{subequations}}
\newcommand{\barr}{\begin{array}}
\newcommand{\earr}{\end{array}}
\newcommand{\nn}{\nonumber \\}

\def\C {{{\cal C}}}
\def\D {{{\cal D}}}

\def\N {{{\cal N}}}

\begin{document}
%%%%%%%%%%%%%%%%%%%%%%%%%%%%%%%%%%%%%%%%%%%%%%%%%%%%%%%%%%%%%%%%%%%%%%%%%%%%%%%%%%%%%%%
\title{Symmetry conserving Coupled Cluster Doubles wave function and the Self-Consistent odd particle number RPA}
%%%%%%%%%%%%%%%%%%%%%%%%%%%%%%%%%%%%%%%%%%%%%%%%%%%%%%%%%%%%%%%%%%%%%%%%%%%%%%%%%%%%%%%
%\author{Mohsen Jema\"{\i}, Peter Schuck}
%%%%%%%%%%%%%%%%%%%%%%%%%%%%%%%%%%%%%%%%%%%%%%%%%%%%%%%%%%%%%%%%%%%%%%%%%%%%%%%%%%%%%%%%

\date{\today}

\author{M. Jema\"i}
\affiliation{Laboratory of Advanced Materials and Quantum Phenomena, Physics Department, FST, El-Manar University, 2092 Tunis, Tunisia
%Laboratoire des mat\'eriaux avanc\'es et ph\'enom\`enes quantiques, FST, Universit\'e Tunis El-Manar, Campus Universitaire Farhat Hached, B.P. n° 94 - ROMMANA, Tunis 1068, Tunisia
}
%\email{ mohcen.jemai@issatm.u-carthage.tn }
\affiliation{ISSATM, Carthage University, Avenue de la R\'epublique P.O. Box 77 - 1054 Amilcar, Tunis, Tunisia.}
\email{mohcen.jemai@issatm.u-carthage.tn}
%mohcen.jemai@gmail.com
\author{P. Schuck}
\affiliation{ Universit\'e Paris-Saclay, CNRS-IN2P3, IJCLab, 91405 Orsay Cedex, France.}
\affiliation{ Universit\'e Grenoble Alpes, CNRS, LPMMC, 38000 Grenoble, France }
%\altaffiliation[Also at the]{Laboratoire des materiaux avanc\'es et ph\'enom\`enes quantique,, D\'epartement de Physique, Facult\'e des Sciences de Tunis, Universit\'e de Tunis El-Manar 
%                          2092 El-Manar, Tunis, Tunisie. }
\email{schuck@ipno.in2p3.fr}
%Peter.Schuck@grenoble.cnrs.fr
%%%%%%%%%%%%%%%%%%%%%%%%%%%%%%%%%%%%%%%%%%%%%%%%%%%%%%%%%%%%%%%%%%%%%%%%%%%%%%%%%%%%%%%
\begin{abstract}
Mixing single and triple fermions an exact killing operator of the Coupled Cluster Doubles (CCD) wave function with good symmetry was found in \cite{Tohy13}. Using these operators with the equation of motion (EOM) method the so-called self-consistent odd particle number random phase approximation (odd-RPA) was set up. Together with the stationarity condition of the two body density matrix it is shown that the killing conditions allow to  reduce the order of correlation functions contained in the matrix elements of the odd-RPA equations to a fully self consistent equation for the single particle occupation numbers. Excellent results for the latter and the ground state energies are obtained in an exactly solvable model from weak to strong couplings. 
\end{abstract}

\pacs{Random phase approximation, Coupled cluster
double wave function, restoration of broken symmetry, 
Lipkin model.}

\maketitle
%%%%%%%%%%%%%%%%%%%%%%%%%%%%%%%%%%%%%%%%%%%%%%%%%%%%%%%%%%%%%%%%%%%%%%%%%%%%%%%%%%%%%%%
\section{Introduction}
\label{Introd}

It is well known that the coupled cluster doubles (CCD) wave function is a powerful many-body ansatz. 
However, it is not easy nor straightforward to perform calculations with the CCD wave function.
The technique most in use \cite{Bla86, Bar07, Pap14} is to project the equations for the ground state energy onto successively more complicated m$p$-m$h$ configurations with m = 1, 2, ... and $p, h$ single particle (s.p.) states above and below the Fermi level, respectively. Often excellent results have been obtained with these methods in various fields of physics (nuclear physics, chemistry, condensed matter, ...) \cite{Bar07,Pap14}. 
However, the method runs into difficulties when the system under consideration undergoes a transition to a spontaneously broken symmetry. A typical example is the transition to superconductivity of electronic systems or to super-fluidity of other Fermi systems like there are nuclear physics or cold atoms in traps. This is particularly relevant for finite systems where considering a definite number of particles can become mandatory. Very recently there have, thus, been attempts to formulate symmetry projected CDD approaches: i) using BCS quasi-particle basis with projection to good particle number \cite{Scu13} ii) an effort has also been undertaken for parity projection in the Lipkin model \cite{Scu17}. Evidently such techniques lead to quite complex equations and to the best of our knowledge particle number projected CCD has not been applied to any realistic system so far.\\
In this paper we will go a different route leading certainly also to a theory of quite some complexity but which presents in our opinion rather interesting aspects. In the recent past it was shown in \cite{Tohy13} that the CCD wave function is the vacuum to exact killing operators mixing single and triple numbers of fermion operators. Taking those operators within an equation of motion (EOM) approach leads to the so-called self-consistent odd particle number RPA (odd-RPA) approach \cite{Tohy13}. The problem is how to evaluate the matrix elements which contain up to three-body correlation functions appearing in this odd-RPA in a consistent way, since those operators which consist in a non-linear transformation of fermions cannot be inverted as it is the case with quasi particle operators obtained, e.g., from a Bogoliubov transformation among fermions. However, as we will show in this paper, there exists a way around. One namely can use the killing condition which mixes, as mentioned, single and triple fermion operators to reduce the order of correlation functions \cite{Virgil}. In applying this method successively, we will achieve that the matrix elements in odd-RPA only contain correlated s.p. occupation numbers. To achieve this, one also has to take advantage in a last step of the stationarity of the two particle density matrix. We demonstrate the excellent performance of this approach in applying it to the Lipkin model. This model is exactly solvable and frequently used to test many body techniques mostly in nuclear physics where it was invented but not only \cite{Lip65, Dus04, Vid04, Rib07, Gco18, Cas08, Pueb15, Colo15, Camp15}. This model contains for instance a phase with a spontaneously broken symmetry mixing even and odd numbers of $ph$ excitations. It is, therefore,  a discrete symmetry (parity), in $ph$ space which is broken in strong coupling. We will see that the results from odd-RPA for s.p. occupation numbers and ground state energies are excellent in the weak and strong coupling limits and still being of very acceptable accuracy in the transition region.\\

The paper is organised as follows. In Sect.\ref{General=Th} we present the general theory and in Sect.\ref{App-Lipkin} we apply it to the Lipkin model. In Sect.\ref{Concl-Persp} we give our conclusions and outline some perspectives. Finally in the Appendix we give some detailed formulas.

\section{General Theory}
\label{General=Th}
\noindent

In this paper we will consider the following CCD wave function

\ba
\vert {\rm Z}\rangle &=& \exp\left({\rm Z}\right)|{\rm HF}\rangle \label{CCD}
\ea
with
\ba
{\rm Z} &=&\frac{1}{2}\sum_{pp'hh'} z_{pp'hh'} K ^\dag_{ph} K^\dag_{p'h'} 
\nn
&=& \frac{1}{2}\sum_{pp'hh'} z_{pp'hh'} P ^\dag_{pp'} P_{hh'} ,
\ea
\noindent
where $\vert {\rm HF}\rangle$  is the Hartree-Fock (HF) Slater determinant and
\ba
K^\dag_{ph}=\beta^\dag_{p} \beta^\dag_{h} 
~~~~~~~~~~~
K_{hp}= \beta_{h} \beta_{p}
\ea
\noindent
with $\beta_h|{\rm HF}\rangle = a^\dag_h|{\rm HF}\rangle = 0$ and $\beta_p|{\rm HF}\rangle = a_p|{\rm HF}\rangle = 0 $ ($a^{\dag}_k, ~a_k$ the fermion operators in the HF basis). The indices "$p,~ p', \ldots $" refer to single particle states 'above' and "$h,~ h', \ldots $" refer to single hole states 'below' the Fermi surface, respectively.\\
The pairing operators are given by
\be
P^{\dag}_{pp'} =\beta^\dag_{p} \beta^\dag_{p'} 
~~~~~~~~~~~
P^\dag_{hh'}= \beta^\dag_{h} \beta^\dag_{h'}
\ee
The amplitudes $z_{pp'hh'} $ must full-fill the condition of the killer operators of the state (\ref{CCD}).

For an odd particle excitation operator, the killers can be  defined as retrieval ($\rho$) mode or addition ($\alpha$) mode, respectively
\ba
q_{\rho} &=& \sum\limits_h x^\rho_h \beta _{h} +\sum_{pp'h'} U^\rho_{pp'h'} \beta^\dag_{p} K^\dag_{p'h'}  
\nn 
q_{\alpha}&=& \sum\limits_p x^\alpha_p \beta _{p} +\sum\limits_{p'hh'} U^\alpha_{p'hh'} K^\dag_{p'h'} \beta^\dag_{h}.
\label{odd-operatorsxU}
\ea

\noindent
The killing conditions $q_{\rho}\vert {\rm Z}\rangle =q_{\alpha}\vert {\rm Z}\rangle =0 $ give
\ba
\sum\limits_{h'}  z_{pp'hh'} x^\rho_{h'} &=& U^\rho_{pp'h} ,~~~~\sum\limits_{p'} x^\alpha_{p'} z_{p'phh'} = U^\alpha_{phh'}.~~
\label{xz=U}
\ea
For the case of pairing we can consider the following killers 

\ba
q_{\rho}  &=&\sum_{h} y^\rho_{h} \beta_{h} +\sum_{pp'h'} V^\rho_{pp'h'}   P^\dag_{pp'}\beta^\dag_{h'}
\nn 
q_{\alpha}&=&\sum_{p}y^\alpha_{p}\beta_{p} +\sum_{p'hh'} V^\alpha_{p'hh'} \beta^\dag_{p'}P^\dag_{hh'}
\label{odd-operatorsyV}
\ea
and the killing conditions give the relations
\ba
\sum\limits_{h'}  z_{pp'hh'} y^\rho_{h'} &=& V^\rho_{pp'h} ,~~~~\sum\limits_{p'} y^\alpha_{p'} z_{p'phh'} = V^\alpha_{phh'}.~~
\label{yz=V}
\ea
In this paper we will, however, not consider the pairing case any further. It will be treated separately in a forthcoming paper.
\noindent
The coefficients $x^\rho,~U^\rho$ will be determined from the minimisation of a sum rule for the average single particle energy
\ba
\lambda_\rho =    \frac{\langle\{q_\rho,[H,q^\dag_\rho]\}\rangle}{\langle\{q_\rho,q^\dag_\rho\}\rangle}
\label{sumRule-odd}
\ea

\noindent
and equivalently for $x^\alpha,~U^\alpha$ with $q^\dag_{\alpha}$.
From the minimisation of equation (\ref{sumRule-odd}), we obtain two coupled equations 
\ba
\sum\limits_{h'} e_{hh'} x^\rho_{h'} + \sum\limits_{pp'h'} \C_{h,pp'h'} U^\rho_{pp'h'} = \lambda_\rho x^\rho_{h} ~~~~~
\\
\sum\limits_{h'} \C^{*}_{pp'h,h'} x^\rho_{h'} + \sum\limits_{p_1p_2h'}\D_{pp'h,p_1p_2h'} U^\rho_{p_1p_2h'} = \lambda_\rho U^\rho_{pp'h}\nonumber
\ea

\noindent
or written as a matrix eigenvalue equation  
\ba
\begin{pmatrix} e  & \C \\ \C^{\dag} & \D \end{pmatrix}  
\begin{pmatrix} x^\rho \\ U^\rho \end{pmatrix}
=\lambda_\rho 
\begin{pmatrix} x^\rho \\ U^\rho \end{pmatrix}
\label{odd_matrix}
\ea

\noindent
with
\ba
e_{hh'} &=& \langle\left\{\beta _{h},\left[H,\beta ^\dagger_{h'}\right]\right\}\rangle 
\nn
\C^{*}_{pp'h',h} &=&\frac{\langle \left\{K _{h'p'} \beta _{p} ,\left[H,\beta ^\dagger_{h} \right] \right\} \rangle }{\sqrt{\N_{pp'h'}}}
\nn
\D_{pp'h,p_1p_2h'} &=&\frac{\langle\left\{K _{hp'}\beta _{p},\left[H,\beta ^\dag_{p_1} K^\dag_{p_2h'} \right] \right\}\rangle}{\sqrt{\N_{pp'h}}\sqrt{\N_{p_1p_2h' }}}
\nn
\N_{pp'h'} &=& \langle\left\{K_{h'p'} \beta_{p}, \beta ^\dag_{p} K^\dag_{p'h'}\right\}\rangle
\label{elemt-matodd} 
\ea
where $\langle \ldots \rangle =\frac{\langle {\rm Z}| \ldots |{\rm Z}\rangle}{\langle {\rm Z}|{\rm Z}\rangle} $ will be used throughout the paper.\\
The Hamiltonian of two-particles in interaction is given by 
\begin{equation}
H =\sum \limits_{kl} \epsilon_{kl}  c^{\dag}_{k} c_{l}
+\frac{1}{4}\sum\limits_{klmn} \bar{v}_{klmn} c^{\dag}_{k} c^{\dag}_{l} c_{n} c_{m}.
  \label{HFermion}
\end{equation}
\noindent
with $\epsilon_{kl}$ represents the matrix of the kinetic energy. The anti-symmetrised matrix elements of the two-body force are given by 
$\bar v_{klk'l'}=\langle kl|v|k'l'\rangle -\langle kl|v|l'k'\rangle $.
A general two-body Hamiltonian in the HF-quasi-particle basis is given by \cite{RS80}
\be
H=E_{HF} +H^{11} +H^{20}+H^{40} +H^{31} +H^{22} \label{Ham2corps}
\ee
The different terms in (\ref{Ham2corps}) are defined as
\ba
E_{HF} &=&\sum\limits_h \epsilon_{hh} +\tfrac{1}{2} \sum\limits_{hh'} \bar{v}_{hh'hh'} ,
\nn
H^{20} &=&\sum\limits_{ph}\biggl( \epsilon_{ph} +\sum\limits_{h'} \bar{v}_{ph'hh'}\biggl) K^\dag_{ph} +{\rm h.c.}
\nn
H^{40} &=&\frac{1}{4} \sum\limits_{pp',hh'} \bar{v}_{pp'hh'} K^\dag_{ph} K^\dag_{p'h'} +{\rm h.c.},
\nn
H^{11} &=&\sum\limits_{pp'} \biggl(\epsilon_{pp'}+\sum\limits_{h}\bar{v}_{php'h}\biggl) S_{pp'}
\nn
&&~-\sum\limits_{hh'} \biggl(\epsilon_{h'h} +\sum\limits_{h_1} \bar{v}_{h'h_1hh_1} \biggl)S_{hh'} 
\nn
H^{31} &=&\frac{1}{2} \sum\limits_{ph} K^\dag_{ph} \biggl(\sum\limits_{p'p_1} \bar{v}_{pp'hp_1} S_{p'p_1}    
\nn
&&~-\sum\limits_{h'h_1} \bar{v}_{ph_1hh'} S_{h'h_1} \biggr) + {\rm h.c.},
\nn
H^{22} &=&\sum\limits_{php'h'} \bar{v}_{ph'hp'} K^\dag_{ph} K _{h'p'}
\nn
&&+\tfrac{1}{4} \sum\limits_{pp'p_1} \bar{v}_{pp_1p'p_1} S_{pp'}
+\tfrac{1}{4} \sum\limits_{hh'h_1} \bar{v}_{hh_1h'h_1} S_{h'h}   
\nn
&&-\tfrac{1}{4} \sum\limits_{pp_1p'p_2} \bar{v}_{pp'p_1p_2} S_{pp_2} S_{p'p_1} 
\nn
&&-\tfrac{1}{4} \sum\limits_{hh'h_1h_2} \bar{v}_{hh'h_1h_2} S_{h_1h'} S_{h_2h} ,
\ea
\noindent
The density operators $S_{ij}$ are given by
\ba
S_{hh'}=\beta^\dag_{h} \beta _{h'}
~~~~~~~~~~~
S_{pp'}=\beta^\dag_{p} \beta _{p'}
\ea
\noindent
We now will proceed to the reduction of the order of correlation functions contained in the matrix elements of (\ref{elemt-matodd}). We start with the following relations
\ba
\beta_{k} \vert {\rm Z}\rangle &=& e^Z \tilde \beta_{k} \vert {\rm HF}\rangle 
\ea
\noindent
whith $\tilde \beta_k = e^{-Z}\beta_ke^{Z}$. Then
\ba
\tilde \beta_{p}  &=&\beta_{p} +\left[\beta_{p}, Z \right] + {\rm zero} 
\nn
&=&\beta_{p} + \sum_{p'hh'} z_{pp'h'h'} K ^\dag_{p'h} \beta ^\dag_{h'}  
\nn
\tilde \beta_{h} &=&\beta_{h} +\left[\beta_{h},Z \right] + {\rm zero} 
\nn
&=&\beta_{h} +\sum_{pp'h'} z_{pp'hh'} K ^\dag_{ph'} \beta ^\dagger_{p'}. 
\ea

\noindent
This yields the following relations
\sba
\ba
\beta_{p} \vert {\rm Z}\rangle &=& 
\sum_{p'hh'} z_{pp'hh'} K^\dag_{p'h} \beta^\dag_{h'}  \vert {\rm Z}\rangle
\\
\beta_{h} \vert {\rm Z}\rangle &=&\sum_{pp'h'} z_{pp'hh'} K^\dag_{ph'} \beta ^\dag_{p'} \vert {\rm Z}\rangle
\ea
\sea
\noindent
which are just variants of the killing conditions. Now multiplying these relations from the left with $\beta_h (\beta_p)$ and using (\ref{xz=U}), we arrive at a reduction of higher powers in $K^\dag $ to lower powers ones 
\sba
\ba
\sum_{p'p_1h_1h'} U^\rho_{pp'h} z_{p_1p_2h_1h'} K ^\dag_{ph} K ^\dagger_{p_2h_1} K ^\dag_{p'h'} \vert {\rm Z}\rangle =
\nn
\delta_{pp_1} x^\rho_{h} K _{hp_1} \vert {\rm Z}\rangle -(U^\rho_{p_1ph} - U^\rho_{pp_1h}) K^\dag_{ph} \vert {\rm Z}\rangle 
\label{reduction1p} 
\\
\sum_{p_1p'h_1h_2} U^\alpha_{ph_1h'} z_{p_1p'hh_2} K^\dag_{ph_1} K^\dag_{p_1h_2} K^\dag_{p'h'} \vert {\rm Z}\rangle  =
\nn
-\delta_{hh'} x^\alpha_{p} K _{hp} \vert {\rm Z}\rangle - (U^\alpha_{phh'} -U^\alpha_{ph'h}) K^\dag_{ph'} \vert {\rm Z}\rangle 
\label{reduction1h}
\ea
\sea

\noindent 
Similarly, we can reduce the even powers of $K^\dag $  
\sba
\ba
\sum_{pp'} U^\rho_{pp'h} K^\dag_{ph'} K^\dag_{p'h} \vert {\rm Z}\rangle &=& x^\rho_{h} S _{h'h} \vert {\rm Z}\rangle
\label{correl-even-powers1p}
\\
\sum_{hh'} U^\alpha_{phh'} K^\dag_{ph} K^\dag _{p'h'}\vert {\rm Z}\rangle &=&- x^\alpha_{p} S _{p'p} \vert {\rm Z}\rangle
\label{correl-even-powers1h}
\ea
\sea

\noindent
From the mean value in the ground state (\ref{CCD}), we find the s.p. occupation numbers 
\ba
n_h &=& \langle \beta^\dag_h \beta_h\rangle =\sum\limits_{\rho} \vert \langle\{\beta^\dag_h ,q_\rho\}\rangle\vert ^2 =\vert x^\rho_h\vert ^2 
\nn
n_p &=& \langle \beta^\dag_p \beta_p\rangle =\sum\limits_{\alpha} \vert \langle\{\beta^\dag_p , q_\alpha\}\rangle\vert ^2 =\vert x^\alpha_p\vert ^2 
\label{density-ni-nm}
\ea

\noindent
and 
\ba
\langle S_{hh'}\rangle &=& \langle \beta^\dag_{h} \beta_{h'}\rangle 
\nn
&=&\sum\limits_{\rho} \langle\{\beta^\dag_{h} ,q_\rho\}\rangle \langle\{q^\dag_\rho, \beta_{h'} ,\}\rangle = x^\rho_{h} (x^\rho_{h'})^* 
\nn
\langle S_{pp'}\rangle &=& \langle \beta^\dag_{p} \beta_{p'}\rangle 
\nn
&=&\sum\limits_{\alpha} \langle\{\beta^\dag_{p} ,q_\alpha\}\rangle \langle\{q^\dag_\alpha , \beta_{p'}\}\rangle = x^\alpha_{p} (x^\alpha_{p'})^* 
\label{density-Sij-Smn}
\ea
\noindent
We note that the mean values in the ground state (\ref{CCD}) of odd powers of $K^\dag $ vanish. But from the eqs. (\ref{correl-even-powers1p}, \ref{correl-even-powers1h}) and (\ref{density-Sij-Smn}), we can calculate all mean values of even powers of $K^\dag $. Let us add the two eqs (\ref{reduction1p}, \ref{reduction1h}), we then can express any correlation functions appearing in the elements of
the matrix (\ref{odd_matrix}) as functions of $\langle S_{pp'}\rangle $, $\langle S_{hh'}\rangle $ and the mean value of the square of these operators in (\ref{density-Sij-Smn}). In order to close the system of equations, we need one further relation. It very naturally is given by demanding that the time derivative of the two body correlation function be zero. It is this stationary condition of the two-body density matrix which gives us a relation between the $\langle S_{kk'}S_{ll'}\rangle $ and $\langle S_{nn'}\rangle $,
\ba
\langle\left [H,K^\dag_{ph} K^\dag_{p'h'}\right]\rangle =\langle\left [H,K_{hp} K_{h'p'}\right]\rangle =0.
\label{stationaryCondGene}
\ea
\noindent
For the explicit form of this commutator, see appendix (\ref{StatCond}).
In order to test our theory, we choose the Lipkin model for  an application.

\section{Application to the Lipkin model}
\label{App-Lipkin}
\noindent
The single-particle space of the Lipkin model consists of two fermion levels, each of which has a N-fold degeneracy see \cite{Lip65,RS80}. The upper (lower) level has the energy of $\frac{e}{2}$ ($-\frac{e}{2}$). The Hamiltonian of the Lipkin model is given by 

\begin{equation}
H=e J_0 -\frac{V}{2} \left(J^2_+ +J^2_-\right)
\label{Hlipkin}
\end{equation}
\noindent
with $e $ the inter-shell spacing, $V$ is the coupling constant, and 
\ba
J_{0}&=&\frac{1}{2}\sum_{m=1}^{N}\left(c_{1m}^{+}c_{1m}-c_{0m}^{+}c_{0m}\right) , 
\nn
J_{+}&=&\sum_{m=1}^{N}c_{1m}^{+}c_{0m},, ~~~~~J_{-}=(J_{+})^{\dagger } 
\label{su2_operators}
\ea
\noindent
with $2J_0 = \hat{n}_1 -\hat{n}_0 $, $\hat{n}_i =\sum c^{\dagger}_{im} c_{im}$ and $N$  the number of particles equivalent to the degeneracies of the shells.\\
The Lipkin model has been derived in nuclear physics and is, as mentioned in the Introduction, exactly solvable and frequently used to test many body approaches. The model is non-trivial and  has a spontaneously discrete (parity) broken symmetry phase. Besides in nuclear physics, it is also considered in other fields of physics, see \cite{Dus04, Vid04, Rib07, Gco18, Cas08, Pueb15, Colo15, Camp15}.

\noindent
To proceed to the odd-RPA approach, we assume as variational ground state the CCD wave function given by the following expression
\begin{equation}
|{\rm Z}\rangle =e^{\frac{z}{2} J_{+} J_{+}}|HF\rangle =e^{Z}|{\rm HF}\rangle
\label{corr-stat}
\end{equation}

\noindent
The ground state (\ref{corr-stat}) is the vacuum for the two killers of normalised retrieval and addition modes, respectively,
\ba
q_{\rho} &=& \frac{1}{N}\sum_m\left[x^\rho_0 c^\dag_{0m} +U^\rho_0\frac{c^\dag_{1m} J_+}{\sqrt{n_{11}}} \right]
\nn 
q_{\alpha}&=&\frac{1}{N}\sum_m\left[x^\alpha_1 c_{1m}  + U^\alpha_1 \frac{J_+ c_{0m}}{\sqrt{n_{11}}} \right]
\label{odd-ansatz22}
\ea
with $n_{11}= \langle\{c^\dag_{1m} J_+, J_- c_{1m} \} \rangle = \langle\{c^\dag_{0m} J_-, J_+ c_{0m} \} \rangle $. So we can verify the normalisation condition as 
\ba
\langle \{q_{\rho},q^\dag_{\rho}\}\rangle &=& |x^\rho_0|^2+|U^\rho_0|^2  =1
\nn
\langle\{q_{\alpha},q^\dag_{\alpha}\}\rangle &=& |x^\alpha_1|^2+|U^\alpha_1|^2 =1
\ea
Let us calculate the transformed single particle operators
\ba
\tilde c^\dag_{0m}  &=& c^\dag_{0m} -z c^\dag_{1m} J_+ 
\nn
\tilde c_{1m}   &=& c_{1m} + z  J_+ c_{0m} ~,
\ea
and  consider the normalised amplitudes  
\be
\tilde{U}^\rho_0=\frac{U^\rho_0}{\sqrt{n_{11}}},~~~~~~~~\tilde{U}^\alpha_1=\frac{U^\alpha_1}{\sqrt{n_{11}}}.
\ee
Then the condition $ q_{\rho} |{\rm Z}\rangle =0$ and $ q_{\alpha} |{\rm Z}\rangle =0$ yields
\ba 
z=\frac{\tilde{U}^\rho_0}{x^\rho_0}=\frac{U^\rho_0}{x^\rho_0\sqrt{n_{11}}}=\frac{\tilde{U}^\alpha_1}{x^\alpha_1}=\frac{U^\alpha_1}{x^\alpha_1\sqrt{n_{11}}}
\ea
We can find an expression of $J_0|{\rm Z}\rangle $ via $c_{0m} q_{\rho} |{\rm Z}\rangle =0 $,
\ba
J_0 |{\rm Z}\rangle =\left(-\tfrac{N}{2}  + z J_+ J_+\right) |{\rm Z}\rangle \label{J0|Z}
\ea

\noindent
and for $J_-|{\rm Z}\rangle $ via $c_{1m} q_{\rho} |{\rm Z}\rangle =0 $,
\ba
J_-|{\rm Z}\rangle = z\left(\tfrac{N}{2} -  J_0 \right) J_+|{\rm Z}\rangle .\label{Jm|Z}
\ea

\noindent
We can use the two equations (\ref{J0|Z}) and (\ref{Jm|Z}) to find an expression for the correlation functions in terms of $\langle J_0\rangle $ and $\langle J^2_0\rangle $ (see more details in the Appendix. \ref{CalCorrFun}). 

It remains to express $\langle J^2_0\rangle $ as a function of $\langle J_0\rangle $. For this, one uses the stationary condition of the two bodies density (\ref{stationaryCondGene}),
\ba
0 &=& \langle\left [H, J^2_-\right ]\rangle  
\nn
&=& 2e\langle J^2_-\rangle -V\langle (2J_0+4J^2_0 -4J_+J_-J_0)\rangle
\ea
Then, we obtain the following expression for $\langle J^2_0\rangle $,
\ba
\langle 4 J^2_0\rangle &=& 2(2N-3)\langle J_0\rangle +2\frac{2V-ez}{Vz}\langle J^2_+\rangle \nn
&&~~~~~-2(N-2)\langle J_+J_-\rangle
\nn
&=&2(2N-3)\langle J_0\rangle +\frac{2V-ez}{Vz^2}(N+2\langle J_0\rangle) 
\nn
&&~~~~~-2(N-2)\langle J_+J_-\rangle
\label{jpjmCDD}
\ea
\noindent
with $\langle J_+J_-\rangle$ given in Appendix B.
Finally all correlation functions are well expressed as a function of $z$ and $\langle J_0\rangle $.
Let us calculate the $\langle J_0\rangle $ using the odd-RPA equations. We consider the conjugate of the killer of the ground state $|Z\rangle $ as odd excitation operator
and we minimize the energy corresponding to these operators (\ref{odd-ansatz22}), 
\ba
\lambda_\rho &=&\frac{\langle\{q_\rho,[H,q^\dag_\rho]\}\rangle}{\langle\{q_\rho,q^\dag_\rho\}\rangle}
~~\mbox{or}~~
\lambda_\alpha = \frac{\langle\{q_\alpha,[H,q^\dag_\alpha]\}\rangle}{\langle\{q_\alpha,q^\dag_\alpha\}\rangle}.
\label{minim-odd}
\ea
We obtain a matrix eigenvalue equation for the two modes with the Hamiltonian matrix
\ba
h_{ij}=\left(\begin{matrix}
h_{00} & h_{01}\\
h_{10} & h_{11}
\end{matrix}\right) 
\ea
and the corresponding secular equation 
\ba
\det(h-\lambda I )= 0
\label{matrixeqlipk}
\ea
\noindent
The normalisation factor $n_{11}$ is given by
\ba
n_{11}&=&\frac{1}{N}\sum_m\langle\{c^\dag_{1m} J_+,  J_- c_{1m}\}\rangle 
\nn
&=&\frac{1}{N}\sum_m\langle\{ J_+ c_{0m}, c^\dag_{0m} J_-\}\rangle 
\nn
&=&\left(1-\tfrac{2}{N}\right)\left(\langle J_+J_-\rangle -\langle J_0\rangle \right) 
+\tfrac{2}{N} \langle J^2_0\rangle 
\ea
\begin{widetext}

\begin{figure}[h]
\includegraphics[width=17cm,height=12cm]{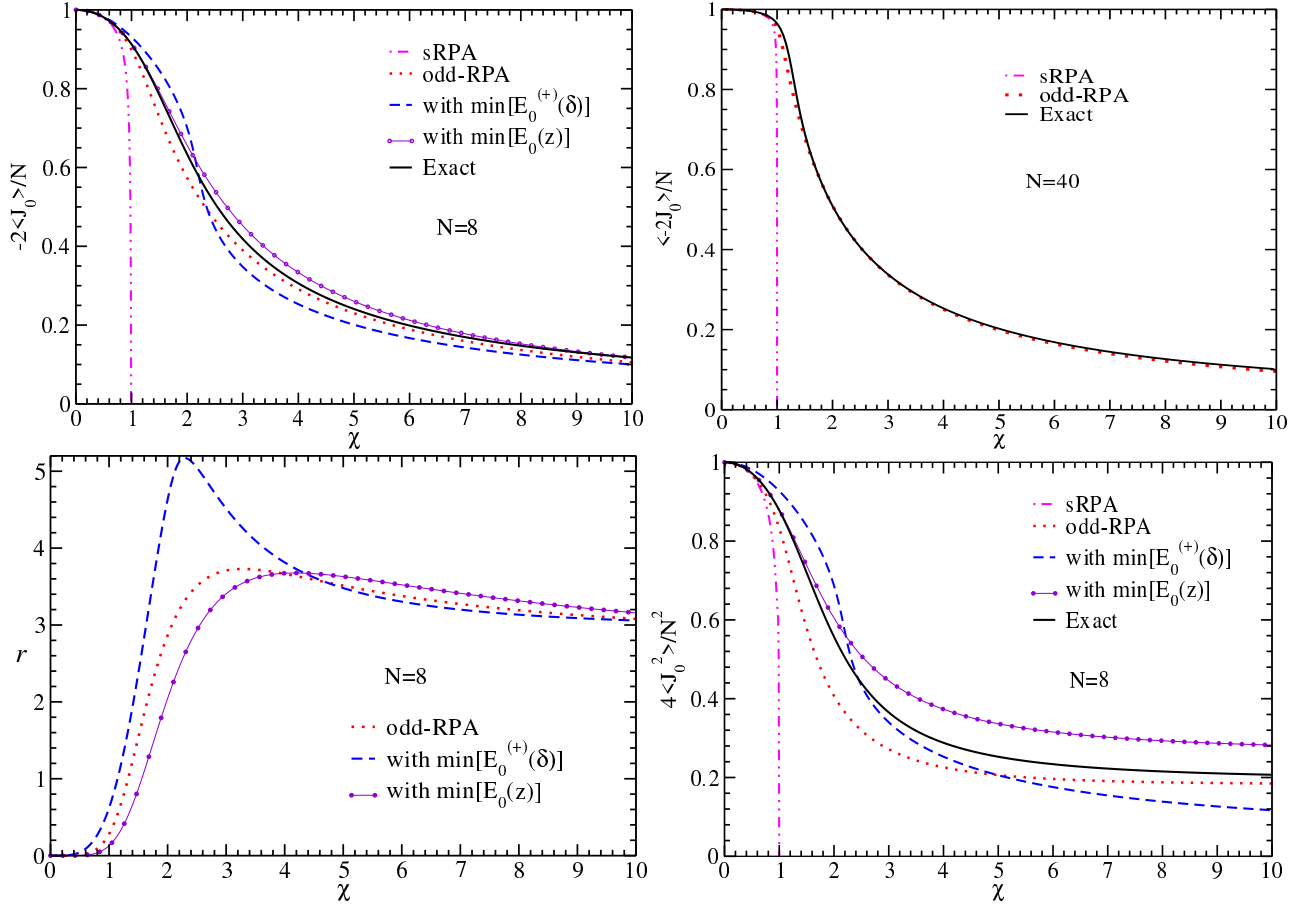}
\caption{Upper left panel: the occupation number difference between upper and lower levels, $\langle -2J_0\rangle $, for $N=8$  with standard RPA (sRPA) (double dot broken line), present odd-RPA (dotted line), projected HF min[$E_0^{(+)}(\delta)$] (broken line), CCD variational wave function min[$E_0(z)$] (continuous line with dots), and exact solution (full line) as function of the intensity of interaction $\chi=\frac{V}{e}(N-1)$. Upper right panel:$\langle -2J_0\rangle $, for $N=40$ with sRPA, odd-RPA, and exact solution. Lower left panel: For $N=8$, percentage error of the correlation energy as $r=100\times \frac{(E_0^{odd-RPA} - E_0^{Exact})}{E_0^{Exact}} $ (dotted line), $r=100\times\frac{(min[E_0(z)]-E_0^{Exact})}{E_0^{Exact} }$ (continuous line with dots) and $r=100\times\frac{(min[E^{(+)}_0(\delta)]-E_0^{Exact})}{E_0^{Exact}} $ (broken line) as function of the intensity of interaction $\chi=\frac{V}{e}(N-1)$. Lower right panel: occupation fluctuation $\langle 4J^2_0\rangle $ for $N=8$ with same ingredients as upper left panel.}
%sRPA, odd-RPA and exact solutions as function of the intensity of interaction $\chi=\frac{V}{e}(N-1)$.}
\label{carre}
\end{figure} 

\begin{figure}
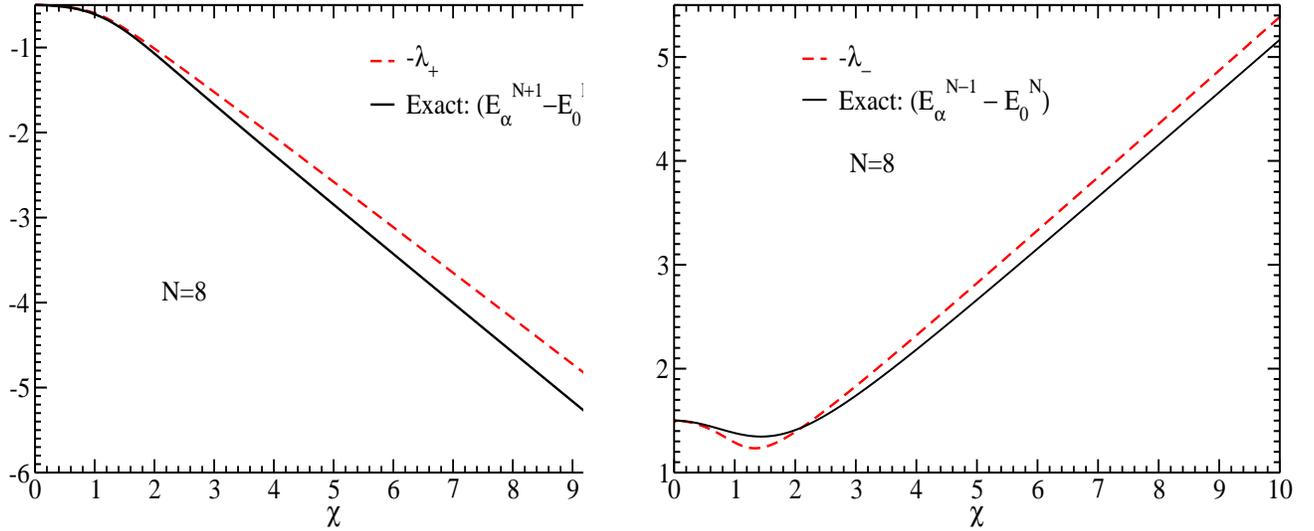

\includegraphics[width=8.5cm,height=7cm]{LpN8.eps}
\includegraphics[width=8.5cm,height=7cm]{LmN8.eps}
\caption{ The eigenvalues $\lambda_+$ and $\lambda_-$ of the odd-RPA matrix compared to the exact values as a function of $\chi$.}
\label{LpLmN10}
\end{figure} 

\end{widetext}
\noindent
For the first Hamiltonian element, we have
\ba
h^\rho_{00} &=&\frac{1}{N}\sum_m \langle\{c^\dagger_{0m},\left[H,c_{0m}\right]\}\rangle = \frac{e}{2} 
\nn
h^\alpha_{00} &=&\frac{1}{N}\sum_m \langle\{c_{1m},\left[H,c^\dag_{1m}\right]\}\rangle = h^\rho_{00}
\ea
\noindent
and for the off diagonal elements  
\ba
h^\rho_{10} &=& h^\rho_{01}=\sum_m  \frac{\langle\{ c^\dagger_{1m}J_+,[H,c_{0m} ]\}\rangle}{N\sqrt{n_{11}}}
= V\sqrt{n_{11}}
\nn
h^\alpha_{10} &=& h^\alpha_{01}=\sum_m\frac{\langle \{ J_+ c_{0m}, [H, c^\dag_{1m} ]\}\rangle}{N\sqrt{n_{11}}} =h^\rho_{10} 
\ea

\noindent
The anti-commutator for $h_{11}$ is given by
\ba
h^\rho_{11}&=&\sum_m\frac{\langle\{c^\dagger_{1m} J_+,[H, J_- c_{1m}]\}\rangle }{N n_{11}} 
\nn
&=&-\frac{3e}{2}-\frac{2V}{Nn_{11} }(N-4)[\langle J^2_+\rangle +\langle J^2_+J_0\rangle ]
\nn
h^\alpha_{11}&=&\sum_m\frac{\langle\{J_+ c_{0m} ,[H, c^\dag_{0m}J_- ]\}\rangle}{N n_{11}}  
=h^\rho_{11}
\ea
\noindent
Then eq.(\ref{matrixeqlipk}) yields
\ba
\lambda _\pm=-\frac{1}{2}\left(e-h_v\right)\pm \frac{1}{2}\sqrt{\left(2e+h_{v}\right)^2 + 4V^2 n_{11}} \label{valpropLipk}
\ea
where %$h_{11}=-\frac{3e}{2}-h_v$ and 
$h_v=\frac{2V}{Nn_{11} }(N-4)[\langle J^2_+\rangle +\langle J^2_+J_0\rangle ]$.
\noindent
So we can calculate
\begin{eqnarray}
n_{0}=\sum_m|\langle c^\dagger_{0m}|\lambda_+\rangle|^2 
= \sum_m|\langle\{c^\dagger_{0m},q_{0,\rho }\}\rangle|^2_{E=\lambda_+}
\end{eqnarray}

\noindent
with $\{c^\dag_{0m},q_{\rho}\} = x^\rho_{0} $. Then, the occupation numbers are given by 
\ba
n_{0}&=& N |x^{+}_0|^2 =N \frac{\lambda _+ -h_{11}}{\lambda_{+} -\lambda _-}
= N\frac{1}{1+z^2 n_{11} }
\ea
\noindent
Thus (with $n_1 = N-n_0$)
\ba
\langle J_0\rangle &=& \frac{N}{2} - n_0 =\frac{N}{2}\frac{z^2n_{11}-1}{z^2n_{11}+1}
\label{2j0Wx0}
\ea
With these relations the odd-RPA equation boils down to a non-linear relation for $z$ which can easily be solved. As can be seen from Fig.\ref{carre}, the results for $\langle J_0\rangle,~ \langle J_0J_0\rangle $, and the correlation energy are excellent for weak to strong coupling. At both ends, the method gives very good results whereas in the transition region the error remains below $3.75\%$. The relative error in the ground state energy for $\chi \rightarrow \infty$ is estimated to be $\sim 3\%$ for $N=8$ and better for higher particle numbers. Also the single particle energies are well reproduced, see Fig.\ref{LpLmN10}. These are very satisfying results. We will give further comments in the next section where we also will give a comparison with two other methods: symmetry projected HF and the direct use of the CCD wave function as a variational one.

\section{Discussions, conclusions and perspectives}
\label{Concl-Persp}
It was known for a certain time that the CCD wave function is killed by well chosen combinations of single and triple fermion operators \cite{Tohy13}. However, because of the non-linear fermion transformation, it remained an open problem how to deal with
these operators. In this paper we showed that there exists a very efficient way how to manage  a calculus with such operators. We showed that the more-body correlation functions appearing in the theory can be reduced to expectation values of the density operator with the help of the killing conditions and the stationarity of the two-body correlation function. The system of (odd-RPA) equations
is then fully closed and calculations for s.p. occupation numbers and ground state energies can be performed. We applied the theory to the Lipkin model with very good success. Indeed occupation numbers and correlation energies become excellent in the weak and remain very good in the strong coupling limits with numbers in between, that is in the transition region, which stay below $4\%$ error. This is very satisfying. One may ask about the reason of this success. To this end, we remark that replacing in eq.(\ref{odd-operatorsxU}) the operators $K^{\dag}_{ph}$ by their expectation values, the non-linear transformation reduces to an ordinary linear HF-transformation among single fermion operators. The killing operators stand , therefore, for some sort of symmetry conserving quantum transformation of fermion
operators. One may also say that the method consists of a symmetry conserving particle-vibration coupling (PVC) approach. In fact we performed calculations with
parity projected HF wave functions (see (blue) broken lines in Fig.\ref{carre}) and also using the CCD wave function as a fully variational one (see (violet) lines with dots). We see that for the energies the latter two approaches are performing about the same as odd-RPA (with projected HF slightly worse) and also for the occupation numbers there is not a significant difference between all approaches.

%As a matter of fact we checked that our results are slightly better but close to the standard variation after projection (VAP) approach of mean field theory. It will be interesting to study the relation of both methods, odd-RPA and VAP, in more detail, for instance in what concerns the numerical implications.\\
In Sect.\ref{General=Th}, we also briefly sketched how to adapt odd-RPA to the pairing problem. This will be a task for the future. A still more ambitious project will be to apply our theory to the case of broken rotational symmetry. However, before, we shall gain more experience with this novel method for simpler cases. Another open problem to be considered in the future is the fact that there exist killers of the CCD wave function which contain an even number of fermion operators \cite{Jem19}. Those operators consist in a slight generalisation of the standard RPA $ph$ operators. Similar procedures as we used here can certainly also be applied for those operators. It shall be very interesting to see how well excitation energies of collective nuclear states are reproduced.

\section*{Acknowledgements}
PS wants to thank Mitsuru Tohyama for past collaboration on odd-RPA. Discussions with Jorge Dukelsky are greatfully acknowledged as well as for suggestions and a carefull reading of the manuscript.

%We are greatful for longstanding collaboration on SCRPA with D. Delion, J. Dukelsky, and M. Tohyama.

\appendix
\section{Stationary condition}
\label{StatCond}
The stationary condition implies that the expectation value of the commutator $[H,K^\dag K^\dag ]$ must be zero. Let us then calculate this commutator with the general Hamiltonian (\ref{Ham2corps}),
\ba
\left[H^{11},K^\dag_{ph}K^\dag_{p'h'}\right]&=& 
 2\sum\limits_{p_1} \varepsilon_{pp_1}K^\dag_{p_1h}K^\dag_{p'h'}
 \nn
&&-2\sum\limits_{h_1} \varepsilon_{hh_1}K^\dag_{ph_1}K^\dag_{p'h'}
\ea
\ba
&&\left[H^{20},K^\dag_{ph}K^\dag_{p'h'}\right]=
 \varepsilon_{ph} K^\dag_{p'h'} 
+\varepsilon_{p'h'} K^\dag_{ph}
\nn
&&~~~~~~~~~~~~~~~~~~~~~~~
-\varepsilon_{ph'} K^\dag_{p'h} -\varepsilon_{p'h} K^\dag_{ph'}
\nn
&&-2K^\dag_{ph} \biggl(
\sum\limits_{h_1} \varepsilon_{p'h_1} S_{h'h_1} +
\sum\limits_{p_1} \varepsilon_{p_1h'} S_{p'p_1}\biggr)
%\nn
%-K^\dag_{p'h'}\biggl(
%\sum\limits_{h_1} \varepsilon_{ph_1} S_{hh_1} +
%\sum\limits_{p_1} \varepsilon_{p_1h} S_{pp_1}\biggr)
\ea
with 
\ba
\varepsilon_{pp'}&=&\epsilon_{pp'}+\sum\limits_{h_1}\bar{v}_{ph_1p'h_1},
\nn
\varepsilon_{h'h}&=&\epsilon_{h'h}+\sum\limits_{h_1}\bar{v}_{h'h_1hh_1},
\nn
\varepsilon_{ph}&=&\epsilon_{ph}+\sum\limits_{h_1}\bar{v}_{ph_1hh_1}.
\ea
\begin{widetext}
\ba
4\left[H^{40},K^\dag_{ph}K^\dag_{p'h'}\right]&=&
-\sum\limits_{p_1h_1}\left(\bar{v}_{p_1h_1p'h} K^0_{ph',h_1p_1} + \bar{v}_{p_1h_1ph'} K^0_{p'h,h_1p_1}\right)
\nn
&&+\sum\limits_{p_1p_2h_1h_2} \bar{v}_{p_1h_1p_2h_2}\biggl(K^0_{ph,h_1p_1} K^0_{p'h',h_2p_2} + K^0_{p'h',h_1p_1} K^0_{ph,h_2p_2}\biggr)
\nn
&&+2\sum\limits_{p_1p_2h_1h_2} \bar{v}_{p_1h_1p_2h_2}\biggl(K^\dag_{ph}K_{h_1p_1}K^0_{p'h',h_2p_2} +K^\dag_{p'h'}K_{h_1p_1}K^0_{ph,h_2p_2}\biggr)
\nn
&&+\sum\limits_{p_1h_1}\biggl (\bar{v}_{p'h_1p_1h'} +\bar{v}_{p_1h'p'h_1}\biggr) K^\dag_{ph}K_{h_1p_1}
  +\sum\limits_{p_1h_1}\biggl (\bar{v}_{p_1hph_1} +\bar{v}_{ph_1p_1h}\biggr) K^\dag_{p'h'}K_{h_1p_1}
  \nn
&&-\sum\limits_{p_1h_1}\biggl (\bar{v}_{p'h_1p_1h} +\bar{v}_{p_1h_1p'h}\biggr) K^\dag_{ph'}K_{h_1p_1}
  -\sum\limits_{p_1h_1}\biggl (\bar{v}_{p_1h_1ph'} +\bar{v}_{ph'p_1h_1}\biggr) K^\dag_{p'h}K_{h_1p_1}
\ea
\ba
\left[H^{22},K^\dag_{ph}K^\dag_{p'h'}\right]&=&-\sum\limits_{p_1h_1}\biggl (\bar{v}_{p_1hh_1p'} K^\dag_{ph'} + \bar{v}_{p_1h'h_1p} K^\dag_{p'h}\biggr) K^\dag_{p_1h_1} 
\nn
&+&\sum\limits_{p_1p_2h_1h_2} \bar{v}_{p_1h_2h_1p_2} K^\dag_{p_1h_1}\biggl (K^\dag_{ph} K^0_{p'h',h_2p_2}+ K^\dag_{p'h'} K^0_{ph,h_2p_2}\biggr)
\nn
&-&\frac{1}{4}\sum\limits_{p_1p_2p_3}\biggl [ (\bar{v}_{p_2p_3pp_1}+\bar{v}_{p_2p_3p_1p})K^\dag_{p'h'}K^\dag_{p_3h}S_{p_2p_1}+ (\bar{v}_{p_3p_2p'p_1}+\bar{v}_{p_2p_3p_1p'})K^\dag_{ph}K^\dag_{p_2h}S_{p_3p_1} \biggr]
\nn
&-&\frac{1}{4}\sum\limits_{p_1p_2}\biggl  [ 
\biggl (\bar{v}_{p_1p_2p'p}  + \bar{v}_{p_2p_1pp'} \biggr ) K^\dag_{p_1h}K^\dag_{p_2h'} 
+\bar{v}_{p_1p_2p'p_2}  K^\dag_{ph}K^\dag_{p_1h'} + \bar{v}_{p_1p_2pp_2}  K^\dag_{p'h'}K^\dag_{p_1h} \biggr]
\nn
&-&\frac{1}{4}\sum\limits_{h_1h_2h_3}\biggl [ \biggl (\bar{v}_{h_3hh_1h_2}+\bar{v}_{hh_3h_1h_2}\biggr ) K^\dag_{p'h'}K^\dag_{ph_1}S_{h_2h_3}+ \biggl (\bar{v}_{h'h_3h_2h_1}+\bar{v}_{h_3h'h_1h_2}\biggr ) K^\dag_{ph}K^\dag_{p'h_1}S_{h_2h_3}\biggr ]
\nn
&-&\frac{1}{4}\sum\limits_{h_1h_2}\biggl [ 
\biggl (\bar{v}_{h'hh_1h_2}  + \bar{v}_{hh'h_2h_1}\biggr) K^\dag_{ph_1}K^\dag_{p'h_2} 
+\bar{v}_{h'h_2h_1h_2} K^\dag_{ph}K^\dag_{p'h_1} +\bar{v}_{hh_2h_1h_2} K^\dag_{p'h'}K^\dag_{ph_1}\biggr ]
\nn
&+&\frac{1}{2}\sum\limits_{p_1p_2} \bar{v}_{p_2p_1pp_1} K^\dag_{p_2h} K^\dag_{p'h'}
+\frac{1}{2}\sum\limits_{h_1h_2} \bar{v}_{hh_1h_2h_1} K^\dag_{ph_2} K^\dag_{p'h'}
\ea
\ba
2\left[H^{31},K^\dag_{ph}K^\dag_{p'h'}\right]&=&K^\dag_{ph} \sum\limits_{p_1h_1} K^\dag_{p_1h_1} \biggl (\sum\limits_{p_2} \bar{v}_{p_1p_2h_1p'} K^\dag_{p_2h'} - \sum\limits_{h_2} \bar{v}_{p_1h'h_1h_2} K^\dag_{p'h_2}\biggr )
\nn
&+& K^\dag_{p'h'} \sum\limits_{p_1h_1}  K^\dag_{p_1h_1}\biggl (\sum\limits_{p_2} \bar{v}_{p_1p_2h_1p} K^\dag_{p_2h} - \sum\limits_{h_2} \bar{v}_{p_1hh_1h_2} K^\dag_{ph_2}\biggr)
\nn
&+& K^\dag_{ph} \sum\limits_{p_1h_1}\biggl (\sum\limits_{p_2} \bar{v}_{p_1p'h_1p_2} K^\dag_{p_2h'} - \sum\limits_{h_2} \bar{v}_{p_1h_2h_1h'} K^\dag_{p'h_2}\biggr ) K_{h_1p_1}
\nn
&+& K^\dag_{p'h'} \sum\limits_{p_1h_1}\biggl (\sum\limits_{p_2} \bar{v}_{p_1ph_1p_2} K^\dag_{p_2h} - \sum\limits_{h_2} \bar{v}_{p_1h_2h_1h} K^\dag_{ph_2}) K_{h_1p_1}
\nn
&+&\biggl (K^\dag_{ph}+K^\dag_{p'h'}\biggl ) \sum\limits_{p_1h_1} K_{h_1p_1}\biggl (\sum\limits_{p_2p_3} \bar{v}_{p_1p_2h_1p_3} S_{p_3p_2} - \sum\limits_{h_2h_3} \bar{v}_{p_1h_3h_1h_2} S_{h_3h_2}\biggr ) 
\nn
&+& \biggl (\sum\limits_{p_2p_3} \bar{v}_{p'p_2hp_3} S_{p_3p_2} - 
     \sum\limits_{h_2h_3} \bar{v}_{p'h_3hh_2} S_{h_3h_2}\biggr )K^\dag_{ph'}
\nn
&+&\biggl (\sum\limits_{p_2p_3} \bar{v}_{pp_2h'p_3} S_{p_3p_2} - 
 \sum\limits_{h_2h_3} \bar{v}_{ph_3h'h_2} S_{h_3h_2}\biggr )K^\dag_{p'h}
\nn
&+&\sum\limits_{p_1h_1}\biggl (\sum\limits_{p_2p_3} \bar{v}_{p_1p_2h_1p_3} S_{p_3p_2} - 
\sum\limits_{h_2h_3} \bar{v}_{p_1h_3h_1h_2} S_{h_3h_2}\biggr )\biggl ( K^\dag_{p'h'}K^0_{ph,h_1p_1}+K^\dag_{ph}K^0_{p'h',h_1p_1}\biggr )
\ea
with
\ba
\left[K_{h_1p_1},K^\dag_{p_2h_2}\right]&=&\delta_{p_1p_2} \delta_{h_1h_2} -\delta_{p_1p_2} S_{h_1h_2} -\delta_{h_1h_2} S_{p_1p_2} 
\nn
&=& K^0_{p_2h_2,h_1p_1}
\ea
\end{widetext}
\noindent
Summing the mean values of the different commutators in the ground state $\vert {\rm Z}\rangle $ for $\langle [H,K^{\dag}_{ph}K^{\dag}_{p'h'}]\rangle =0$ yields a relation between $\langle S_{kk'}S_{ll'}\rangle $ and $\langle S_{nn'}\rangle $.

\section{ Calculation of correlation functions in the Lipkin model}
\label{CalCorrFun}
All correlation functions can be expressed in terms  of $z$, $\langle J_0\rangle $ and $\langle J^2_0\rangle $. We also have
\ba
n_0 &=& \tfrac{N}{2}-\langle J_0\rangle , ~~~~~
n_1 = \tfrac{N}{2}+\langle J_0\rangle
\ea

\noindent
From the first equation (\ref{J0|Z}), we  find
\ba
z\langle J_{+}^{2}\rangle =\tfrac{N}{2}+\langle J_{0}\rangle \label{Jp2}
\ea
Multiplying (\ref{J0|Z}) by $J^2_+$, gives
\ba  
\langle J^2_+ J_{0}\rangle = -\tfrac{N}{2} \langle J^2_+ \rangle +z\langle J_{+}^{4}\rangle  
\ea
which yields
\ba 
2z\langle J_{+}^{4}\rangle = N \langle J^2_+\rangle +2\langle J^2_+ J_{0}\rangle  \label{Jp4-1}
\ea
Multiplying the second equation (\ref{Jm|Z}) by $J_+$, we can write
\ba
\langle J_+ J_{-} \rangle = z(N-1)\langle J^2_{+}\rangle -z^{2} \langle J_{+}^{4}\rangle  
\ea
which leads to
\ba
z^{2}\langle J_{+}^{4}\rangle &=&(N-1)z \langle J^2_{+}\rangle -\langle J_+ J_{-}\rangle \label{Jp4-2}
\ea
Then, with (\ref{Jp4-1}) and (\ref{Jp4-2}), we find
\ba
z\langle J^2_+ J_{0}\rangle =(\tfrac{N}{2} -1)z\langle J^2_{+}\rangle - \langle J_+ J_{-}\rangle 
\label{Jp2J0}
\ea

\noindent
We multiply (\ref{J0|Z}) from the left by $J_0$,
\ba
J^2_0|Z\rangle &=&-\tfrac{N}{2}J_0|Z\rangle +z J_0J^2_+|Z\rangle
\nn
&=&-\tfrac{N}{2}J_0|Z\rangle +2z J^2_+|Z\rangle +z J^2_+J_0|Z\rangle
\label{last}
\ea
Replacing  the last term of (\ref{last}) by its mean value $z\langle J^2_+J_0\rangle$   and  also $z\langle J^2_+\rangle $ by its mean value and using (\ref{Jp2}), we find the Casimir relation
\ba
J_{+}J_{-}|Z\rangle =\left(\tfrac{1}{4}N(N+2)-J^2_0  + J_{0}\right)|Z\rangle
\ea

\noindent
In summary, the correlation functions are given by
\ba
2z\langle J^2_+\rangle &=& N +2\langle J_0\rangle
\nn
\langle J_{+}J_{-}\rangle &=& \frac{1}{4}N(N+2)-\langle J^2_0 \rangle + \langle J_0\rangle 
\nn
z\langle J^2_+ J_0\rangle &=&\frac{1}{2}(N-2)z\langle J^2_+\rangle -\langle J_{+}J_{-}\rangle
\nn
z\langle J_+J_-J_0\rangle &=&\frac{N}{2}(N+2)z +\frac{1}{2}(N-6)z\langle J_{+}J_{-}\rangle 
\nn
&&~ -(N-4)z\langle J_0\rangle -\langle J^2_+\rangle
\ea
what allows to solve the equation for $\langle J_0\rangle$.

\bibliographystyle{refer}
\bibliography{articles}

\end{document}